\documentclass[twocolumn]{article}

\usepackage{graphicx,url,color,simplemargins}
\usepackage{amsmath, amssymb}

\newcommand{\postrev}[1]{{#1}}

\definecolor{light-gray}{gray}{0.95}

\settopmargin{2cm}
\setbottommargin{2cm}
\setleftmargin{1.5cm}
\setrightmargin{1.5cm}

\begin{document}

\title{Explicit tracking of uncertainty increases the power of quantitative rule-of-thumb reasoning in cell biology}
\date{}
\author{Iain G. Johnston\,$^{1}$, Benjamin C. Rickett\,$^{1}$, and Nick S. Jones\,$^{1,}$\footnote{to whom correspondence should be addressed} \\ \footnotesize $^{1}$Department of Mathematics, Imperial College London, Huxley Building, Queen's Gate, London SW7 2AZ, UK.}

\maketitle

\begin{abstract}
`Back-of-the-envelope' or `rule-of-thumb' calculations involving rough estimates of quantities play a central scientific role in developing intuition about the structure and behaviour of physical systems, for example in so-called `Fermi problems' in the physical sciences. Such calculations can be used to powerfully and quantitatively reason about biological systems, particularly at the interface between physics and biology. However, substantial uncertainties are often associated with values in cell biology, and performing calculations without taking this uncertainty into account may limit the extent to which results can be interpreted for a given problem. We present a means to facilitate such calculations where uncertainties are explicitly tracked through the line of reasoning, and introduce a `probabilistic calculator' called Caladis, a web tool freely available at \url{www.caladis.org}, designed to perform this tracking. This approach allows users to perform more statistically robust calculations in cell biology despite having uncertain values, and to identify which quantities need to be measured more precisely in order to make confident statements, facilitating efficient experimental design. We illustrate the use of our tool for tracking uncertainty in several example biological calculations, showing that the results yield powerful and interpretable statistics on the quantities of interest. We also demonstrate that the outcomes of calculations may differ from point estimates when uncertainty is accurately tracked. An integral link between Caladis and the Bionumbers repository of biological quantities further facilitates the straightforward location, selection, and use of a wealth of experimental data in cell biological calculations.
\end{abstract}

\section*{Introduction}
Rule-of-thumb, or back-of-the-envelope, calculations are of great utility across the sciences, allowing estimates of quantities to be obtained while gleaning intuition about the important numerical features of a system. In physics, the paradigm of the `Fermi problem' has been used for decades to develop intuition about the structure and behaviour of systems by employing reasonable approximations, order-of-magnitude estimates, dimensional analysis and clearly stated assumptions. The use of the napkin (often more readily available than an envelope in modern caf\'{e}s and conferences) as a medium to perform rough calculations and gain understanding of a system given limited experimental information is well known in the physical sciences and has recently gained popular attention \cite{weinstein2009guesstimation}. Recent mathematical approaches to complex problems in wider scientific fields have employed these back-of-the-envelope approaches, including `bioestimates' in physical biology \cite{phillips2009physical} and cell biology \cite{milowebsite} and the popular `street-fighting mathematics' for use through the sciences \cite{mahajan2010street}.

\begin{figure}
\includegraphics[width=8cm]{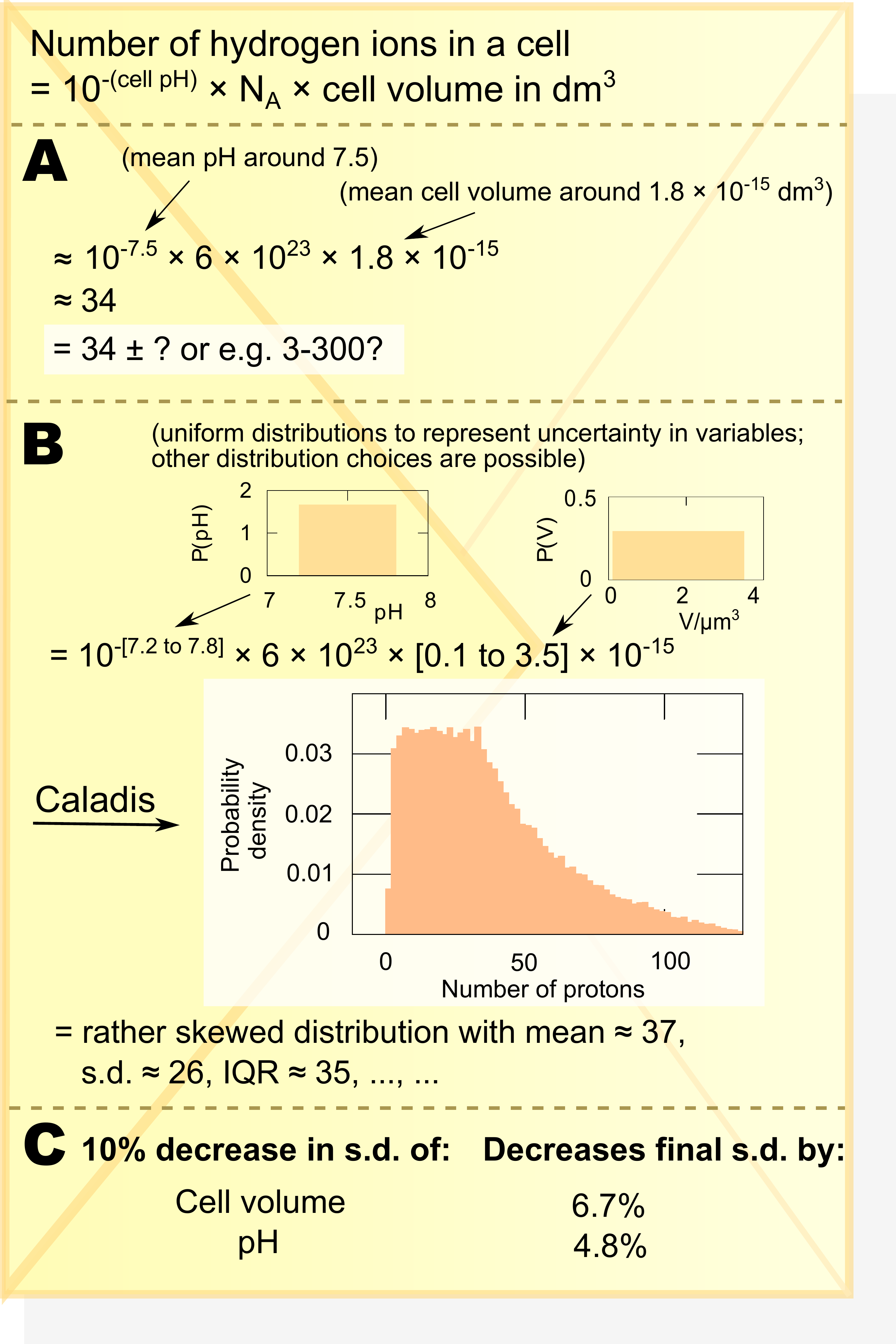}
\caption{\textbf{An example back-of-the-envelope calculation.} Technology is currently unable to measure the number of protons in a cell, so we estimate this number from measured quantities. (A) An estimate without uncertainty, combining rough estimates of pH and cell volume to obtain a guess for the number of protons. In this example, mean values are chosen to match the means of known measurements, but no associated uncertainty is analysed. (B) An estimate using Caladis to explicitly account for uncertainties in the measured quantities and reporting rigorous explanatory statistics about the final quantity, using uniform distributions to represent the uncertainty in the variables involved. Other representative distributions are possible and can be analysed using our approach (see Results). (C) Caladis also finds that in this example calculation, more of the final uncertainty arises from uncertainty in cell volume than pH: refining volume estimates is slightly preferred as the optimal experimental strategy to lower overall uncertainty.}
\label{fig1}
\end{figure}

However, these calculations do not currently have as central a role in cell biology as they do in the physical sciences, despite receiving substantial recent attention as powerful tools for reasoning in quantitative biology \cite{phillips2009feeling, moran2010snapshot}, and being facilitated by quantitative resources like the excellent Bionumbers database \cite{milo2010bionumbers}. One reason for this absence is that many of the quantities involved in cell biology are either intrinsically highly variable or have large measurement errors. Calculations which do not take these uncertainties into account (yielding a mean value estimate with no associated uncertainties), though powerful in their own right, may only represent part of the full story (Fig.~\ref{fig1}A).

In some back-of-the-envelope circumstances, accuracy may be maintained without the explicit tracking of uncertainties. An example of this is in calculations involving the multiplication of several terms, each of which may be reasonably assumed to be normally distributed with similar coefficients of variation. In such a calculation, the logarithm of the error in an estimate scales with the square root of the number of terms in the calculation. However, quantitative cell biology often involves distributions that cannot be assumed to be normally distributed, and calculations more general than simple multiplications of terms. In these circumstances, where individual uncertainties can differ between terms and may be over many orders of magnitude, the risk of inaccuracy associated with calculations without uncertainty is increased. If uncertainties are included in such calculations, it is often through standard propagation-of-uncertainty approaches \cite{regan2004equivalence}, which typically track a limited number of distribution moments and can hence fail to accurately represent the distribution of the final result for non-normal distributions. Of course, the process of performing rough calculations and obtaining estimated answers is immensely valuable in its own right, for the reasons discussed above. In order to complement this powerful process of `Fermi reasoning' in biology, we here suggest an complementary form of `envelope' allowing for calculations including uncertain quantities.

\section*{Methods}
\subsection*{Explicitly tracking uncertainty in cell biological calculations}
We propose an approach to biological rule-of-thumb calculations involving uncertain quantities that does not solely rely on point estimates of quantities of interest. Instead, our approach involves treating every quantity in a rule-of-thumb calculation which has an associated uncertainty as a probability distribution describing this uncertainty. The following iterative process is then performed. In each iteration, a sampled value is taken from each distribution of interest in the calculation. The value of the complete calculation is computed given this set of samples. This process is iterated many times to build up a distribution of values describing the output of the calculation. This output distribution then provides an interpretable and statistically rigorous answer to the rule-of-thumb question. We present this approach as a complement to, and not a substitute for, the valuable process of Fermi estimation, and stress again the value of `just having a go with the numbers'.

We emphasise that our approach, calculation of quantities using samples from distributions rather than point estimates, can be used to obtain interpretable results in cases where we do not have access to the full set of original measurements. This situation is likely to apply, for example, when using summarised results from previous independent experiments. In this case, our method can be viewed as a generalisation of the resampling approaches that may be used if we had full access to the original data, such as bootstrapping or jackknifing \cite{wasserman2004all}.

In addition to providing a statistically rigorous answer to rule-of-thumb questions in cell biology, this approach can also be used to facilitate efficient design of experiments to reduce uncertainty in a given quantity. In the picture of calculations performed using probability distributions, this goal can be accomplished using a simple variant of a sensitivity analysis approach. Consider artificially decreasing the variance of each distribution in a calculation one-by-one. Decreasing the variance of each individual distribution will lead to a decrease in the overall variance of the output distribution, and the magnitudes of these induced overall decreases can be recorded. The quantity with the most power to decrease overall variance in the calculation output can then be identified, and its value further refined through experiment. Conceptually, this approach resembles performing a sensitivity analysis on the variance of the solution distribution with respect to the variances of individual input distributions.

An important point to consider when attempting to quantify uncertainty in scientific calculations is the source and meaning of `uncertainty'. A degree of measurement error may be associated with an experimental protocol, causing uncertainty in the resulting value due to \emph{imprecision}. Alternatively, a given physical or biological quantity may exhibit genuine \emph{variability} independent of the measurement process, in that its value fluctuates or changes with time and/or other controlling factors. The degree to which calculations involving uncertain quantities are interpretable is contingent on the types of uncertainty involved (see Discussion).

\subsection*{Caladis: A probabilistic calculator for biology}
We introduce a web-based calculator called Caladis (\emph{cal}culate \emph{a} \emph{dis}tribution), available for free use at \url{www.caladis.org}, with its open source code also available for download. Caladis, in addition to computing with constant quantities and standard mathematical operators and functions, naturally incorporates probability distributions as fundamental calculation elements, yielding as its output a probability distribution over the final answer. As described above, this probabilistic calculation approach allows uncertainties to be tracked throughout a calculation, providing a wealth of output data and allowing a complete view of the statistical details of the output of a probabilistic calculation (Fig.~\ref{fig1}B) and further information about the sources of uncertainty (Fig.~\ref{fig1}C; see later).

We underline that our web tool requires no knowledge of computer programming and no access to mathematical software tools, and, in addition to functioning on desktop and laptop browsers, is compatible with a range of hand held devices. Our aim in designing this tool is to facilitate fast and easy calculations involving uncertain biological quantities for users including those who lack the background or software to produce their own machinery for performing such calculations. The ability of our site to function on mobile devices makes it a plausible substitute for the well-known napkin over coffee or a conference dinner, facilitating informal but rigorous rough estimates of quantities as new ideas emerge.

\begin{figure*}
\includegraphics[width=18cm]{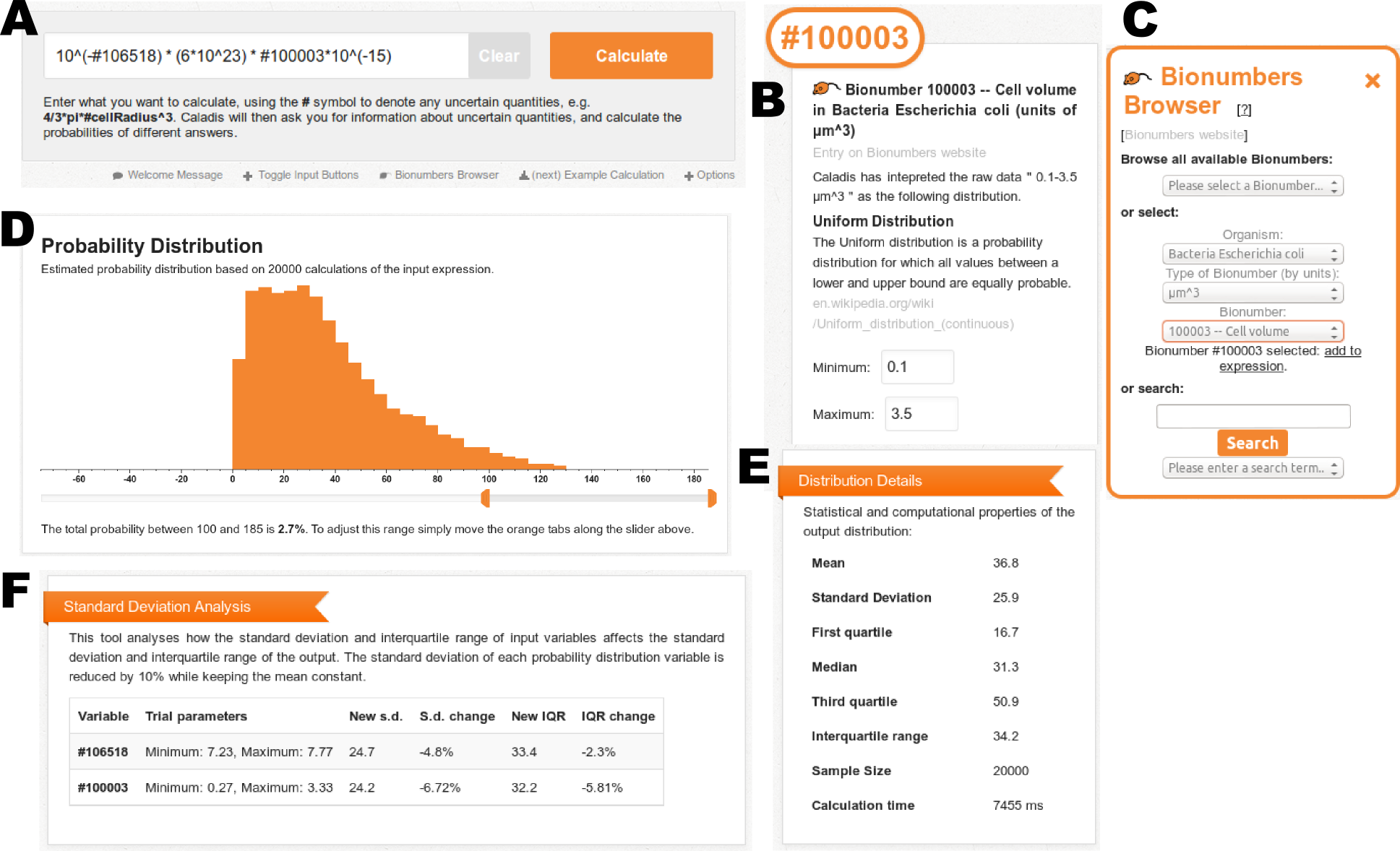}
\caption{\textbf{Elements of Caladis interface.} (A) The expression input box: a user enters a calculation here, providing any required information about any distributions (for example, perhaps specifying that a certain distribution is uniform between $0$ and $1$, or normal with mean $1$ and standard deviation $0.1$). (B) Each probability distribution in the input expression must then be characterised, either through the user's entry of appropriate parameters, or (as depicted) through the automatic recognition of a Bionumber. (C) The Bionumbers Browser allows the identification, selection, and inclusion of values from the Bionumbers database. (D) The resultant distribution for the calculation is then displayed, along with summary statistics of the distribution (E) and (optionally) standard deviation analysis (F) assessing the sensitivity of overall variance with respect to the variance of individual elements. This illustration involves, as an example calculation, the proton number calculation discussed in the Results section.}
\label{fig2}
\end{figure*}

Caladis presents the user with a field (Fig.~\ref{fig2}A) to input calculation expressions, which may involve probability distributions identified with a prepended \# symbol (\# functions as a sigil denoting a distribution, e.g. \texttt{4/3*pi*\#cellRadiusDist\^{}3}). For every probability distribution found in the input expression, Caladis prompts the user to choose a distribution type, and appropriate parameters to describe that distribution (for example, perhaps specifying that \texttt{\#cellRadiusDist} is a uniform distribution between $1 \mu$m and $1.5 \mu$m), or, in the case of Bionumbers (see below), automatically populates the distribution details with the appropriate parameters (Fig.~\ref{fig2}B). Users may also use a built-in browser to input distributions corresponding to recorded quantities from biological experiments (Fig.~\ref{fig2}C; see `Bionumbers' below). The user may then choose to `Calculate' the expression, whereupon Caladis computes a probability distribution describing the final answer using the above approach, sampling many times from each distribution the user entered to build up a set of samples from the resultant distribution, which is then displayed graphically (Fig.~\ref{fig2}D). This interface includes a tool to estimate the probability density between two given values, user-controlled display of the probability density in each bin, summary statistics of the distribution (Fig.~\ref{fig2}E), results from the optional standard deviation analysis (Fig.~\ref{fig2}F), and a URL which serves as a permanent link to that calculation. This collection of output statistics and graphics allows a complete overview of the probabilistic result of the user's calculation.

Caladis also facilitates the aforementioned efficient design of experimental strategies, through consideration of the contributions of different quantities to the overall uncertainty in a calculation. The user has the option of performing a `standard deviation analysis' for common types of input distribution in the web interface. In this analysis, the standard deviation of each input distribution of this type is artificially reduced by 10\%, and the resulting effect on the standard deviation and IQR of the resultant distribution is recorded (Fig.~\ref{fig2}F). The input variable with the most power to refine the overall output estimate can then straightforwardly be identified.

\subsection*{Bionumbers}
We have embedded the data provided by the Bionumbers repository \cite{milo2010bionumbers} within Caladis. Bionumbers contains a huge range of biological measurements, spanning scales from microscopic chemical reaction rates and cellular concentrations to ecosystem- and planet-wide statistics of biological populations. Our link to the database allows us to perform powerful rule-of-thumb biophysical and cell-biological calculations with Bionumbers \cite{phillips2009feeling} while tracking uncertainties in order to estimate the ranges of the final answer. 

Within our web tool, the Bionumbers database is parsed to obtain, for each Bionumber, a corresponding probability distribution, units, and a URL to the source data. Probability distributions are assigned based on the format of the source data and according to a user-defined protocol (see Appendix). The units of each value are automatically obtained from the database. Users may then use a variety of approaches to identify and select Bionumbers for use in a probabilistic calculation, and the corresponding probability distributions are automatically included as calculation elements (see Appendix).

\section*{Results}

\subsection*{Problems with reasoning with mean values in nonlinear contexts} 

We first illustrate how reasoning using only mean estimates may lead to incorrect results in calculations. Consider two measured quantities $X$ and $Y$, perhaps corresponding to the abundance of two different types of entity in a population. We are interested in the proportion of $X$ in the population $P = X / (X+Y)$.

Say we have the information that the measured quantities follow log-normal distributions, with $X$ having mean $m_X = 0.1$ and standard deviation (of the log-normal distribution itself, as opposed to the underlying normal distribution) $s_X = 0.1$, and $Y$ having mean $m_Y = 0.9$ and standard deviation $s_Y = 0.9$. In this artifical example, estimating the expected proportion of $X$ in the population from the means alone would give $\hat{P} = m_X / (m_X + m_Y) = 0.1$. However, accurately tracking uncertainty in this calculation produces the counterintuitive result that $\mathbb{E}(P) \simeq 0.144$, rather more than the population proportion estimated from mean values (Fig.~\ref{fig4}A).

This illustration contrasts with the cases where a calculation is straightforwardly additively or multiplicatively separable. In such cases, the fact that functions $f(X)$ and $g(Y)$ of independent random variables $X$ and $Y$ are themselves independent leads to the results $\mathbb{E}(f(X) g(Y)) = \mathbb{E}(f(X)) \mathbb{E}(g(Y))$ and $\mathbb{E}(f(X)+ g(Y)) = \mathbb{E}(f(X))+ \mathbb{E}(g(Y))$, implying that calculations based on the individual means of $X$ and $Y$ will accurately estimate the overall mean. The error in the mean-based estimate $\hat{P}$ in our example arises from the structure of the expression used to calculate the population proportion: the fraction cannot be separated into independent functions of $X$ and $Y$. Generally in such inseparable cases, calculations based solely on mean values may not provide correct estimators. In such cases, explicitly tracking uncertainty not only provides a powerful characterisation of the uncertainty in the final answer but also guarantees that such errors in the mean outcome are not made.

Next, we give two example calculations from the `Bionumber of the month' website \cite{milowebsite} to illustrate the process of explicitly tracking uncertainties in cell biological calculations with Bionumbers. The details of the Bionumber distributions used are shown in the Appendix.

\subsection*{The number of hydrogen ions in a cell} 

Given measurement of the pH and volume $V$ of a system, the number of hydrogen ions present in the system can be deduced as $n = 10^{-\mbox{\footnotesize pH}} N_A V$, where $N_A \simeq 6 \times 10^{-23}$ is Avogadro's number. In the Dec 2011 entry of Ref. \cite{milowebsite}, measurements of pH and cell volume are used to estimate that an \emph{E. coli} cell contains around 60 hydrogen ions. Using Caladis' Bionumbers browser to search for `cell volume' and `cytoplasm pH' identifies Bionumbers \#100003 (\emph{E. coli} cell volume) and \#106518 (\emph{E. coli} pH). These values appear in Bionumbers as (\#100003) `$0.1-3.5\,\mu \mbox{m}^3$, interpreted as $U(0.1,3.5)\,\mu\mbox{m}^{3}$; and (\#106518) `$7.2$ to $7.8$', interpreted as $U(7.2,7.8)$. It is possible to interpret these results in terms of different probability distributions -- a facility supported by Caladis (see Appendix). For example, the quantity `$0.1-3.5\,\mu \mbox{m}^3$' could be interpreted as a log-normal distribution with $0.1\,\mu \mbox{m}^3$ and $3.5\,\mu \mbox{m}^3$ as $\pm 1 \sigma$ points of the distribution. However, in this specific example, we use a uniform distribution, as the corresponding log-normal distribution exhibits extremely high variance with a range over more than an order of magnitude, which does not intuitively match the expected distribution of cell sizes in a population. Additionally, analytic results for the distribution of exponentially growing, dividing cells suggest a quadratic distribution that bears a stronger resemblance to the uniform than the log-normal picture \cite{rausenberger2008quantifying}. The ability to explore these different interpretations, and quantitatively debate the properties of each, are valuable scientific processes which our approach facilitates.

We can automatically access these Bionumbers and their associated uncertainties in Caladis, then calculate the above equation while tracking uncertainties (this calculation forms the example used illustratively in Fig. \ref{fig1}B). We find that the resultant distribution (see Fig.~\ref{fig4}B) easily spans an order of magnitude, with $14\%$ of the density less than 10 protons and $3\%$ more than 100 protons (statistics straightforwardly found using Caladis' interface). Use of standard deviation analysis suggests that more of this uncertainty originates from the spread of cell volumes. We now have a mean estimate around 37 and a full characterisation of the uncertainty associated with this answer, allowing a quantified degree of confidence to be associated with our reasoning.

\subsection*{Diffusion times in cells} 

In the Mar 2010 entry of \cite{milowebsite}, the characteristic timescales for diffusion through cells of various sizes are explored, using the expression $t = x^2 / 6D$, where $x$ is the length scale of diffusion and $D$ the diffusion constant of the species of interest. Ref. \cite{milowebsite} uses a rough estimate of the diffusion constant for GFP in \emph{E. coli} and order-of-magnitude reasoning to obtain an estimate of 10ms to traverse an r.m.s. distance of 1$\mu$m. 

\begin{figure}
\centering
\includegraphics[width=8cm]{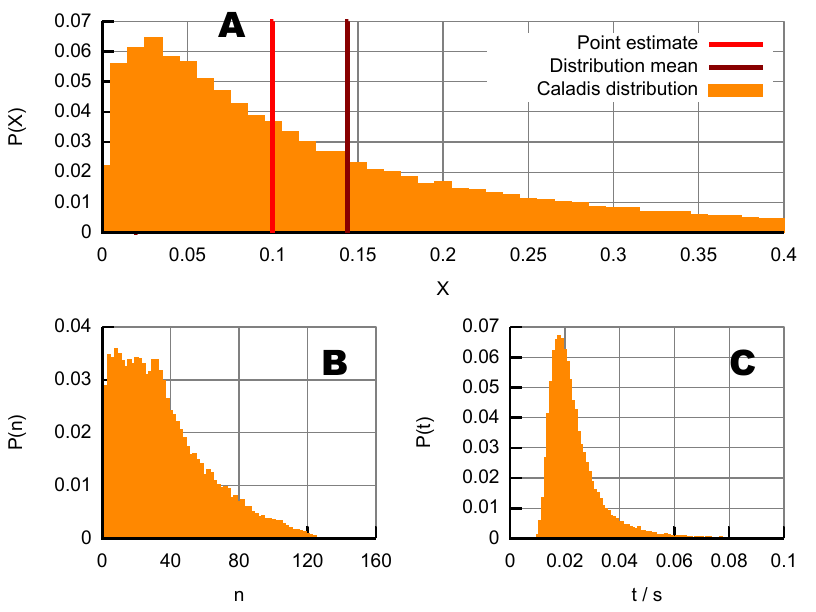}
\caption{\textbf{Point estimates and biological distributions.} (A) The distribution resulting from the illustrative $X / (X+Y)$ calculation in the text. The value obtained by considering mean estimates alone differs from the mean of the true distribution, which is heavily skewed, highlighting the importance of explicitly tracking uncertainty. (B, C) Estimates, using data from biological experiments via the Bionumbers database and tracking uncertainties, of (B) the number of protons in an \emph{E. coli} cell and (C) time for GFP to diffuse $1 \mu$m in \emph{E. coli}. All distributions are direct outputs from Caladis.}
\label{fig4}
\end{figure}

Using Caladis' Bionumbers browser to search for `diffusion rate' identifies Bionumber \#100193 (diffusion rate in \emph{E. coli}), recorded as `$7.7 \pm 2.5\,\mu \mbox{m}^2 \mbox{s}^{-1}$' and interpreted as $N(7.7, 2.5)\, \mbox{m}^2 \mbox{s}^{-1}$. We follow the calculation in Ref. \cite{milowebsite} by including this Bionumber in the above equation, using $x = 1\mu$m, and performing the probabilistic calculation of $t$ in Caladis, tracking uncertainties. We observe that the resultant distribution (see Fig.~\ref{fig4}C) is highly skewed, with an apparent coefficient of variation (the ratio of the standard deviation to the mean, illustrating the spread of the distribution) around 2.4. This example, where a probability distribution appears in the denominator of an expression for a quantity of interest, illustrates how the resultant uncertainty can behave unintuitively when variables are combined even in relatively simple ways. Calculation of a resultant distribution provides a more robust method in these circumstances than traditional propagation-of-uncertainty approaches, and construction of a full probability distribution for the output of a calculation allows interpretation of details like skewness which are missed by a stimple estimate of the standard deviation alone.

\section*{Discussion}
We have described an approach for performing rule-of-thumb calculations in biophysics and cell biology while incorporating the considerable uncertainty often involved in such biological contexts. This approach, which does not rely solely on point estimates of relevant quantities, allows the treatment and interpretation of the uncertainty involved in such calculations, increasing their trustworthiness and their power to assist intuitive reasoning. In addition, it may be used to optimise experimental design, by identifying measurements with the greatest power to refine knowledge of the overall quantity of interest.

To facilitate the straightforward use of this approach, both at a computer and on mobile devices, we have introduced Caladis, an online tool for performing calculations involving probability distributions, available for free use and with its source code open and available to download. Caladis has a particular link to rule-of-thumb calculations with Bionumbers in cell biology, and we have illustrated its use in deriving distributions of quantities of biophysical and cell biological interest. We note that, in employing these calculations in a scientific context, it is important to note that tracking uncertainties in calculations is only useful if the underlying model is appropriately trusted: hygienic treatment of errors is a separate consideration from picking the right model for the world. It is unlikely that the model probability distributions employed in our approach (and many other analyses) represents the perfect description of a quantity arising in the real world; however, we hope that our approach, with the broad range of distributions supported by Caladis, provides a means of reasonably estimating a wide range of real quantities. As we highlight above, the discussion of appropriate models for uncertainty, and their quick quantitative comparison, is a scientifically beneficial feature facilitated by our approach.

Back-of-the-envelope calculations (though used throughout history) have become increasingly popular recently as tools for developing quantitative reasoning and intuition \cite{weinstein2009guesstimation, mahajan2010street, phillips2009feeling}. Despite this increase in popularity, their use is not yet as prevalent in biology as in the physical sciences. We hope that this tool provides support for, and may increase trust in, the use of back-of-the-envelope calculations in quantitative cell biology (and across the biosciences) by exposing the role of uncertainties. We have shown that in some cases (for example, in calculating proportions), failing to track uncertainties can lead to rough guesses that do not represent the full truth of the calculation. 

In our work with biological calculations, we have found that Caladis plays a useful role in quality control for rule of thumb reasoning: after having made an approximate estimate on a real napkin, it is helpful to check whether the biological question at hand remains adequately answered if variability/imprecision is appropriately accommodated. We further note the reverse possibility: rather than serving as a sanity check for our envelope calculations, Caladis can help create optimism in our estimates. For example, in settings where the uncertainty of some calculation elements is known to be very substantial, it might be the case that the final distribution of the estimated quantity is, in fact, sufficient for scientific advance. As discussed, an uncertainty appended to an estimated quantity needs to be treated with care (since it can depend on distribution choice) but it can serve as a partial certificate for the relevance of the estimate. We suggest that researchers may present links to their calculations within Caladis, so that readers are then free to use their prior beliefs to modify the component distributions (if, for example, a reader is less confident about a variable than the author) to see if the conclusions are still robust. 

As mentioned previously, the interpretation of calculations tracking uncertainty is contingent on the source of the uncertainty in the elements of the calculation, which may arise from imprecision (for example, measurement errors associated with an experimental protocol) or variability (the natural fluctuations intrinsic to a system of interest). Care must be taken in the interpretation of the resultant distribution depending on the sources of uncertainty in the calculation. For example, consider a quantity $X$ which is subject to natural variability, stationary but fluctuating with time, and which has been characterised by a distribution involving a finite number $N$ of measurements of $X$ at different times. If we are interested in the behaviour of $X$ over an infinitesimally small time window, it makes sense to draw from this distribution of $X$, as this distribution represents plausible states of the system. If we are interested in the time-averaged behaviour of $X$, we may instead consider the distribution of $\hat{\mathbb{E}}(X)$, an estimate of the mean of $X$. $\mathbb{E}(X)$ is a single number about which we are uncertain: the distribution of $\hat{\mathbb{E}}(X)$ derived from our measurements will have a finite width (the standard error on the mean, dependent on $N$), corresponding to imprecision rather than natural variability. Mixing uncertainties due to imprecision with those due to variability may lead to results which are not trivial to interpret. We underline the importance of transparency in the meaning of a probabilistic calculation to avoid misinterpretation -- in the above example, it should be explicitly stated whether a calculation involves (variable) single instances of a measurement ($X$) or (imprecise) time-averaged behaviour ($\mathbb{E}(X)$).

The process of sampling from distributions describing individual quantities, performing a calculation using these samples, and building a final distribution is akin to several methodologies of use in Bayesian statistics \cite{andrieu2003introduction}. The difference between our approach and Bayesian sampling approaches is that after establishing our distributions we condition on no further data, instead assuming that the individual distributions (which could be pictured as priors) already contain all information on the likelihood of individual values. In this sense, the Bayesian interpretation of our approach is not as a method for extracting posteriors from priors given data; but is instead a method for performing calculations with priors with no new data, thus constructing new prior distributions over more complicated quantities.

\section*{Acknowledgements}

The authors thank Roslyn Lavery for help with web hosting and development, Ron Milo for helpful comments on the manuscript, and the valuable comments of three anonymous reviewers. IGJ acknowledges funding from the UK MRC. NSJ acknowledges grants EP/K503733/1 and BBD0201901.

\bibliographystyle{unsrt}
\bibliography{caladisrefs}

\section*{Appendix -- Technical Details}
This text describes various technical details of the web interface for Caladis, our probabilistic calculator, at \url{www.caladis.org}.

\textbf{Distributions and sampling.} Available distributions in Caladis are normal, uniform, discrete uniform, log-normal, binomial, Poisson, beta, exponential, gamma, and geometric. For each distribution identified in the input expression, Caladis uses Monte Carlo sampling to sample the resultant distribution: each iteration, random samples are drawn from each characterised input distribution and the value of the input expression is calculated and recorded to build up the resultant distribution. \postrev{The user may determine the number of iterations to employ. Resultant} distributions for which the summary statistics have not converged are identified and a warning message encouraging the use of robust statistics (median, IQR) or more samples is displayed.

\textbf{Options.} Caladis users can select the number of Monte Carlo samples, the angle unit (degrees or radians) and binning methods (Freedman-Diaconis, Scott, or Sturges approaches). Additionally, users can elect whether to perform standard deviation analysis, and how various values from the Bionumbers repository are interpreted (see below).

\textbf{Bionumber selection.} Bionumber IDs may be directly entered (e.g. \texttt{\#100001}), or found with a built-in browser (Fig. 2C). This browser enables a user to identify a Bionumber for use in calculations using one of three approaches. Firstly, a given Bionumber may be selected directly from a full listing of all available experimental data. Secondly, a user may navigate through the set of organisms for which Bionumbers are available, and through the types of value present for each organism (classified by the units with which Bionumbers are associated, so that, for example, length scales may be distinguished from reaction rates). Thirdly, a user may search the descriptions of all Bionumbers for terms of interest, then select from the available search results. Upon identification of a Bionumber of interest, the user may automatically enter that Bionumber (with its associated experimental uncertainty) into their calculation. Upon entry, a Bionumber's distribution will automatically be assigned to the input expression (see below).

\textbf{Bionumber distributions.} \postrev{Caladis assigns distributions to Bionumbers by the format of their associated range information, according to a protocol dictated by the user. Data presented as `$x\,\mbox{to}\,y$', `$x\,-\,y$', or similar, may be interpreted as $U(x,y)$, a uniform distribution between $x$ and $y$; or $\exp( N( m, s))$, a log-normal distribution with parameters chosen such that $x$ and $y$ are $\pm1 \sigma$ points of the distribution. This log-normal interpretation is accomplished through the mapping $m = (\ln a + \ln b)/2; s = m - \ln a$, so that the mean and standard deviation of the resultant log-normal distribution are $\mu = \exp (m + s^2 / 2)$ and $\sigma = \sqrt{ (\exp(s^2) - 1) \exp(2 m + s^2)}$. Data presented as `$x \pm y$' may be interpreted as either normal or log-normal with mean $x$ and standard deviation $y$. Hence, for example, Bionumber \#100001 (the cell length of \emph{E. coli}) is listed as `1.94 to 2.72\,$\mu$m', so we may interpret it (in addition to its qualitative details) as $U(1.94,2.72)$ in units of $\mu$m. In the case of Bionumbers with no associated information regarding uncertainty, Caladis automatically assigns a normal distribution with a coefficient of variation of 0.5 (which the user can change manually).}

\textbf{Documentation.} The online documentation at \url{www.caladis.org/tutorial} contains extensive information on the available mathematical functions and operators, syntax, and details of the optional choices. Several example calculations, illustrating syntax and descriptions of probability distributions, are available on the Caladis input screen, along with options regarding the mathematical details of the calculation.

\end{document}